# Molecular simulations have boosted knowledge of CRISPR/Cas9: A Review

Angana Ray[1*] and Rosa Di Felice[1,2,3*]

[1]Department of Physics and Astronomy, University of Southern California, Los Angeles, CA 90089, USA

[2]Department of Biological Sciences, Quantitative and Computational Biology sector, University of Southern California, Los Angeles, CA 90089, USA

[3]CNR Institute of Nanosciences, Via Campi 213/A, 41125 Modena, Italy

**Abstract**

Genome editing allows scientists to change an organism's DNA. One promising genome editing protocol, already validated in living organisms, is based on clustered regularly interspaced short palindromic repeats (CRISPR)/Cas protein-nucleic acid complexes. When the CRISPR/Cas approach was first demonstrated in 2012, its advantages with respect to previously available techniques, such as zinc-finger nucleases (ZFNs) and transcription activator-like effector nucleases (TALENs), immediately got attention and the method has seen a surge of experimental and computational investigations since then. However, the molecular mechanisms involved in target DNA recognition and cleavage are still not completely resolved and need further attention. The large size and complex nature of CRISPR/Cas9 complexes has been a challenge for computational studies, but some seed results exist and are illuminating on the cleavage activity. In this short review, we present recent progress in studying CRISPR/Cas9 systems by molecular dynamics simulations with coarse-grained and atomistic descriptions, including enhanced sampling.

**Introduction**

During the Second International Summit on Human Genome Editing in Hong Kong in November 2018, He Jiankui of the Southern University of Science and Technology (SUST) in Shenzhen, China, announced that a set of twin girls were brought into the world after applying the CRISPR/Cas9 genome editing technique. The girls were born to an HIV-negative mother and an HIV-positive father. He Jiankui reported that the mother was impregnated with embryos modified by CRISPR/Cas9 facilitated knockdown of chemokine receptor (CCR5), making the embryos resistant to HIV infection (1). This is a clear evidence of the huge potential impact of genome editing in human life.

Nearly two decades ago, the paper titled 'Initial sequencing and analysis of the human genome' (2) was published and considered to be a major milestone in the international Human Genome Project. Ever since, human genetics has become the focus of biomedical research and therapeutics. By definition, a genome is an organism's complete set of DNA, which is composed of DNA bases in stacked pairs. The process of accurately changing the DNA bases at predetermined locations is genome editing. The ability to intentionally precisely modify the genetic code is highly valuable and it is the objective of intense research efforts. There currently exist different plausible molecular techniques, of which CRISPR/Cas is the most promising one.

In this short review, after tracing the history of the CRISPR/Cas technique and describing the structure of the molecular complexes, we focus on computational studies of the system, which nicely complement experimental work in the field. Cas is the short name for "CRISPR-associated protein"; there are different such proteins and Cas9, CRISPR-associated protein 9, is one of them.

**Historical Background**

Genome editing techniques like transcription activator-like effector nucleases (TALENs), zinc-finger nucleases (ZFNs) and CRISPR/Cas have already enabled researchers to make precise modifications at the nucleotide level. In ZFNs and TALENs, the DNA sequence is recognized by a protein: thus, different protein mutants need to be engineered for different target sequences. In CRISPR/Cas, instead, there is no need for lengthy protein engineering, because the target DNA sequence is recognized by a complementary RNA sequence bound to an endonuclease protein. The CRISPR/Cas method offers a number of other advantages over TALENs and ZFNs: it is easily programmable, inexpensive and efficient. The CRISPR/Cas timeline **(Figure 1)** is intriguing. It includes results starting from 1987, when the CRISPR cluster repeats were reported for the first time (3). However, it was much later, in August 2012, that Emmanuelle Charpentier, in collaboration with Jennifer Doudna at the University of California Berkeley, demonstrated the use of CRISPR/Cas9 to target and cleave five specific genomic sites of the gene encoding the green fluorescent protein (GFP) (4): they generalized their conclusions to potentially any target DNA sequence. Further significant progress was achieved slightly after, in February 2013, when Feng

Zhang of the Broad Institute of MIT and Harvard in Cambridge, MA, reported the use of CRISPR/Cas9 for editing genomes of cultured mouse cells and human cells (5). In 2014 the first crystal structures of CRISPR/Cas9 (6, 7) complexes were made available in the Protein Data Bank. CRISPR/Cas is now a very fast-moving field of research, with more than 17000 reports available in 2018.

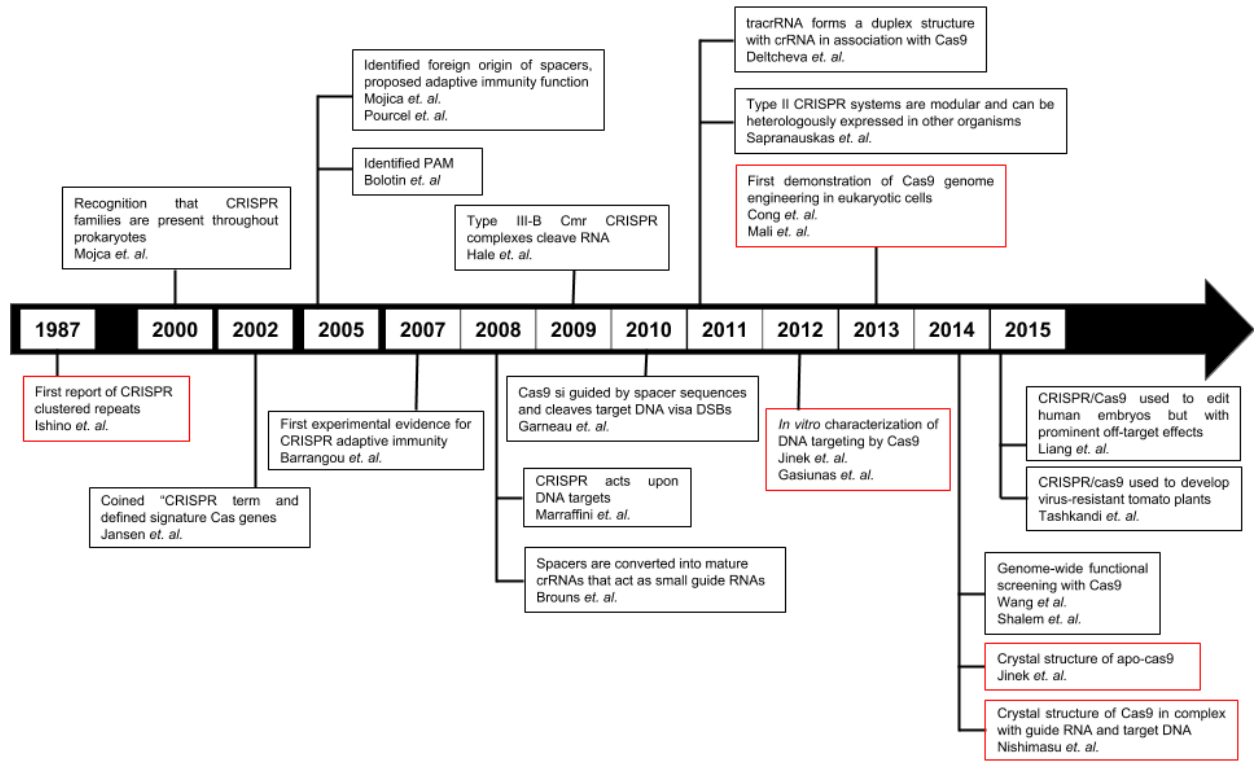

Figure 1: CRISPR/Cas timeline until 2015, with focus on the endonuclease Cas9. The red-colored boxes highlight the most representative events in the history of CRISPR/Cas.

**The Natural Origin of the CRISPR/Cas technique**

Viruses have been a common threat to the survival of bacteria and archaea. As a result of this battle between predator and prey, an array of countermeasures has been adopted by the host organism. The CRISPR system is one such highly adaptive and inheritable bacterial defense mechanism (8) **(Figure 2)**. When a bacterium is infected by a virus and overcomes the infection, it archives fragments of the viral genome, as a memory of the infection. The viral sequences are integrated into the bacterial DNA and intercalated by short repeating elements. The modified bacterial DNA is translated into single strands of RNA that, anchored to suitable proteins, scan new infecting viruses, which are immediately neutralized if the previous infection is recognized through DNA-RNA complementarity. This CRISPR/Cas technique requires the presence of a set of CRISPR-associated (*cas*) genes (9) and acts in three stages: (i) adaptation, e.g., insertion of new spacers in CRISPR locus, (ii) expression, e.g., transcription of CRISPR RNA (crRNA) and (iii) interference, e.g., recognition and destruction of target DNA sequences (9, 10).

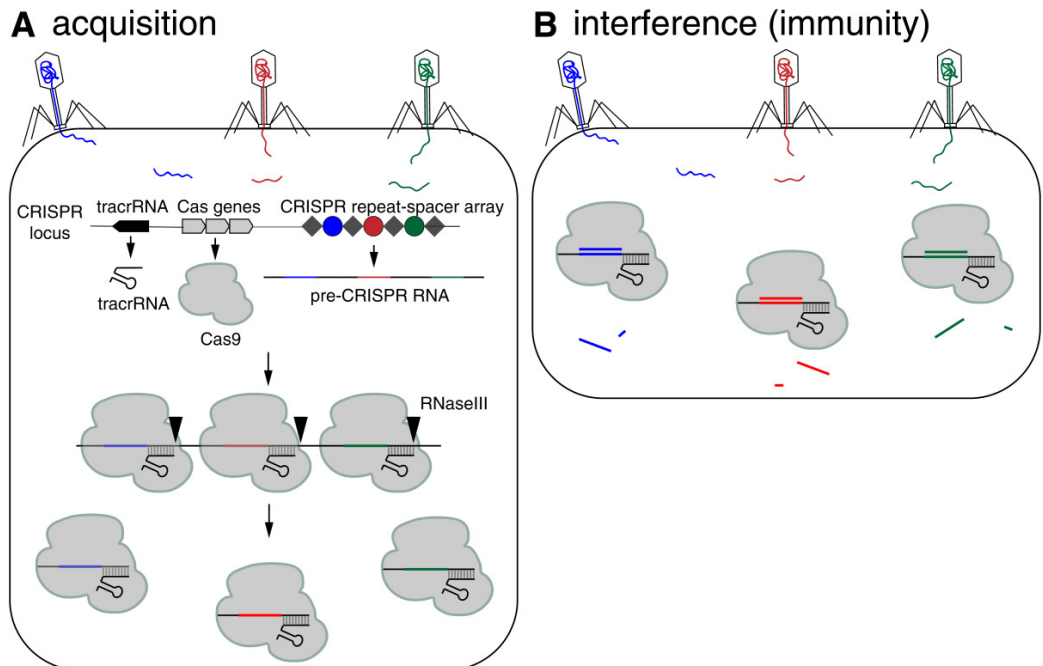

Figure 2: CRISPR/Cas9-mediated acquired immunity in prokaryotes. During the acquisition phase (A), cellular invaders such as phage viruses inject nucleic acid sequences into the host cell. After infection, novel DNA sequences from the cellular invaders are incorporated into the host CRIPSPR locus as spacers (colored circles) flanked by repeat sequences (gray diamonds). As a result, when the CRISPR locus is transcribed, the pre-CRISPR RNAs (crRNAs) encode the newly acquired protospacer sequences. The pre-crRNA is cleaved to produce individual crRNAs that will associate with Cas proteins. The Cas protein utilizes the crRNAs as guides to silence foreign DNA that matches the crRNA sequence (B, interference phase). As a result, the second time a bacterium encounters the same foreign DNA, the crRNA/Cas9 complex is able to identify and silence the DNA. From Ref (11) with permission.

**CRISPR/Cas Classification**

The absence of a universal *cas* gene and its fast evolution (12) has made classification of CRISPR/Cas systems difficult. Hence, multi-level classification is adopted, wherein the CRISPR/Cas systems are broadly classified into two classes, and further subdivided into six types and 33 subtypes (13–15). The difference between the two classes is that the Class 1 systems have multi-subunit effector complexes and in Class 2 systems all functions are performed by a single protein effector module (16). Class 1 includes type I, type III and type IV systems. Class 2 includes the well-known type II systems (including Cas9 and Cpf1), and much more rare type V and type VI systems (14). The classification criteria for classes and types are: a fundamental difference in the organization of the effector modules between two classes; unique signature genes for each of the types. The classification of subtypes is more complex: for certain subtypes, the signature genes are readily defined, while for other subtypes the signature genes are defined through comparison of conserved genes and locus organization (17).

**CRISPR/Cas9 Structure**

In this review we focus on the most popular Class 2 CRISPR/Cas system with the CRISPR-associated protein 9 (Cas9) derived from type II-A CRISPR, obtained from *Streptococcus pyogenes – sp*Cas9. The CRISPR/Cas9 system is a complex composed of single guide RNA (sgRNA) and a 160kDa DNA endonuclease enzyme, Cas9 (18), which cuts each strand of double-stranded DNA at a specific location, through its nuclease domains.

The sgRNA-bound Cas9 endonuclease binds to a double-stranded DNA (dsDNA) upon site-specific recognition of a short trinucleotide Protospacer Adjacent Motif (PAM) within the DNA (19). Thereafter, the target DNA (tDNA) strand, which is complementary to the first 20 nucleotides (-nt) of the sgRNA, forms an RNA-DNA hybrid duplex displacing the non-target DNA (ntDNA) strand from the dsDNA (20). The sgRNA **(Figure 3)** is composed of two noncoding RNA components fused together: a CRISPR RNA (crRNA) that confers target specificity to Cas9 and a transactivating CRISPR RNA (tracrRNA) that can bind to Cas9 (21, 22). The crRNA contains the 20-nt long 'spacer' or 'guide' sequence at the 5'-end that forms the RNA-DNA hybrid, and a 'repeat' sequence at the 3'-end that forms a duplex with the tracrRNA (18, 23, 24). Since the crRNA cannot bind to Cas9 alone, it complexes with the tracrRNA and the resultant RNA:RNA duplex fits into Cas9 (22, 25). A sgRNA can be synthetically generated or obtained *in vitro* or *in vivo* from a DNA template.

The Cas9 endonuclease **(Figure 4)** has two lobes: recognition (REC) lobe and nuclease (NUC) lobe, connected by an Arginine-rich linker. The REC (residues 56-718) and NUC (residues 1-55 and 719-1368) lobes are responsible for association with the sgRNA (7) and cleavage of the DNA, respectively. The NUC lobe consists of different domains: RuvC (residues 1-55, 719-765, and 919-1099), HNH (residues 780-906), and PAM-interacting (PI) (residues 1100-1368) (7, 26–28). The RNA-DNA hybrid duplex, with a negatively charged backbone, is accommodated at the positively charged groove formed between REC and NUC lobes (7). The unwound ntDNA (u-ntDNA) strand is proposed to be hosted, before cleavage, at the HNH/RuvC boundary, through stabilizing electrostatic interactions between the negatively charged backbone of u-ntDNA and the positively charged amino acids of the HNH and RuvC domains (29). The HNH and RuvC domains perform site-specific cleavages of the tDNA and ntDNA strands (30), respectively, resulting in a double-strand break (DSB), thereby inducing the host DNA repair mechanisms (20).

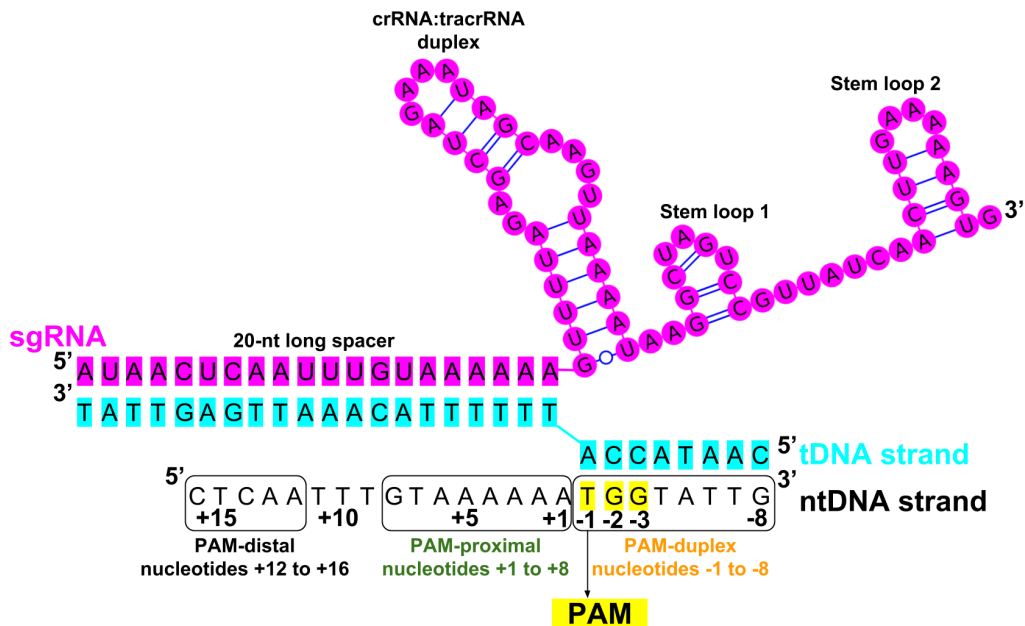

Figure 3: Scheme of the nucleic acid sequence/folding in a CRISPR/Cas9 ternary complex (protein + RNA + DNA) after the Cas9-RNA complex binds to a target DNA. The u-ntDNA strand is absent or short in available crystal structures, including PDB ID 4UN3 (27). In this image, it is elongated according to an electrostatic interaction hypothesis (31). The PAM is highlighted in yellow and numbering of u-ntDNA strand is shown with nucleotides +1 to +16 upstream of PAM. The entire ntDNA strand can be conveniently divided into three segments: PAM duplex, nucleotides -1 to -8; PAM-proximal, nucleotides +1 to +8; PAM-distal, nucleotides +12 to +16.

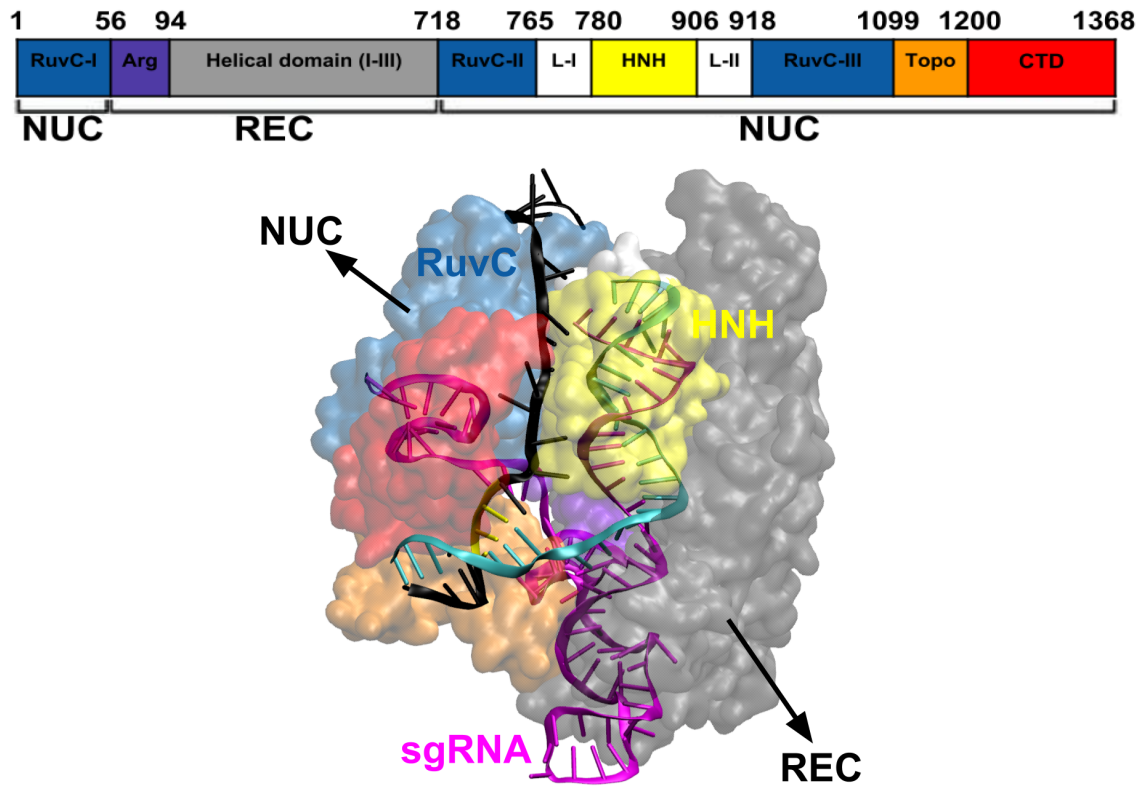

Figure 4: Structural representation of the Cas9 endonuclease bound to RNA and DNA, with the different protein domains represented in different colors, from PDB ID 4UN3 (27) and with the elongated u-ntDNA as in Tangprasertchai *et. al*. (31). The protein is represented with transparent surfaces, the nucleic acids are represented as cartoons, using the visualization software VMD (32).

**Molecular Mechanism of CRISPR/Cas9**

Crystal structures of *Sp*Cas9-sgRNA bound to dsDNA (27, 28) have aided the understanding of the molecular mechanism of Cas9-catalyzed DNA cleavage. Site-specific DNA recognition and cleavage is achieved after Cas9, in complex with guide RNA (crRNA-tracrRNA complex or sgRNA), binds to a target DNA sequence. Thus, as the first step of genome editing by CRISPR/Cas9, the Cas9 protein makes extensive interactions with the sgRNA. Specifically, the crRNA:tracrRNA duplex and stem loop 1 of the sgRNA **(Figure 3)** are crucial for Cas9-RNA complex formation (33); stem loops 2 and 3 may play a role in increasing catalytic efficiency *in vivo* (5, 7, 33, 34). The Cas9-RNA complex then searches for the complementary target DNA sequence, by probing for the PAM (19), highlighted in yellow in **Figure 3**. The PAM sequence is protein-dependent; for *Sp*Cas9, it is 5'-NGG-3', where N can be any DNA base (A, T, G, C). The PAM sequence must be found adjacent to the cleavage site in the ntDNA strand. The GG dinucleotide of the PAM in the ntDNA strand is read out *via* major groove interactions with conserved Arginine residues (R1333 and R1335) of the C-terminal domain (CTD in **Figure 4**) of Cas9, also called PAM-interacting (PI) domain. The PI domain of Cas9 also makes contacts with the minor groove of the PAM-containing duplex. The residues S1136 and K1107 interact with nucleotides of the ntDNA and tDNA strands, respectively, through water-mediated H-bonding. Thus, PAM binding is mainly governed by major groove base recognition, but minor groove interactions between the PAM and Cas9 orient the tDNA strand for base pairing with sgRNA (27). After unwinding of the target DNA duplex, the tDNA strand is stabilized by a "phosphate lock loop", wherein the phosphate group immediately upstream of the PAM is stabilized by amide groups of residues K1107-E1108-S1109 (27).

In a nutshell, local DNA melting is triggered at the PAM-adjacent nucleation site, followed by formation of an "R-loop structure" by the RNA-DNA hybrid duplex and the u-ntDNA strand. Complementary base pairing between the 20-nt long spacer RNA sequence and the tDNA strand is necessary for such target binding specificity and cleavage (18). Upon PAM recognition and subsequent R-loop formation, the Cas9 enzyme is activated for DNA cleavage. The HNH and RuvC domains cleave the tDNA and ntDNA strands, respectively, at a specific site 3 base pairs upstream of the NGG PAM sequence, producing a predominantly blunt-ended DSB (18). In 2016, Zuo and Liu argued, on the basis of MD trajectories, that Cas9-catalyzed DNA cleavage can lead to 1-bp staggered ends, as opposed to blunt ends (35); while the possibility of obtaining staggered ends by Cas9 cleavage is still questionable and needs further evidence, such a cleavage product would be conducive to better homologous recombination strategies.

Another relevant aspect of Cas9 action is that the protein cleaves the target DNA duplex in the presence of divalent metal ions, as HNH and RuvC are $Mg^{2+}$-dependent nuclease domains (36, 37). The RuvC domain has a typical RNase-H fold structure containing four functionally essential residues, D10, E762, H983, and D986, which require a two-metal-ion catalysis mechanism for editing (36–38). The HNH domain, on the other hand, has a ββα-metal fold with three catalytic residues, D839, H840 and N863, consistent with a one-metal-ion cleavage mechanism (36–38). During MD simulations of the CRISPR/Cas9 system in the presence of $Mg^{2+}$ ions, the tDNA and ntDNA strands stabilize near the HNH and RuvC catalytic sites, respectively (30). However, currently available force fields for magnesium ions have limitations, which often result in artifacts of the coordination sphere of the ion (39). This has been a difficulty in obtaining the mechanistic details of how the catalytic $Mg^{2+}$ ions meditate the concerted DNA strand cleavage by two domains of Cas9. However, this issue has been overcome by employing density functional theory (DFT) based quantum-mechanics/molecular-mechanics (QM/MM) hybrid quantum/classical simulations (40), which are presented later in this review.

**MD Simulations of CRISPR/Cas9 complexes**

The large size and complexity of CRISPR/Cas systems are challenging elements for computational studies. The first MD simulations of CRISPR complexes were reported as recently as 2015 and addressed the CRISPR/Csy4 complex. In this work, Estarellas *et. al.* (41) used classical all-atom MD simulations in explicit solvent to understand the basic structure and dynamics of the binary RNA/Csy4 complex, not including the DNA component. The work identified potential catalytic rearrangements of the RNA/Csy4 complex during the simulation time. However, the authors also pointed out that the conclusions from MD trajectories could be affected by the limitations of available simulation techniques. The limitations were highlighted in the same paper by Estarellas and co-workers (41), in order to stimulate in future studies careful analysis and a better understanding of the CRISPR/Cas9 system by MD simulations. Important MD limitations are connected to the constraints on accessible time scales, the resolution of available crystal structures and the accuracy of available force fields; the latter is particularly crucial in molecular systems that include an RNA component (41).

The time-scale limit of MD simulations can be overcome, e.g., by using special hardware (Anton by D.W. Shaw Research) (42) or by adopting coarse-grained modeling methods and simplified energy functions. In 2016, Wenjun Zheng performed coarse-grained MD simulations of a ternary CRISPR/Cas9 complex containing the RNA, DNA, and protein (43). The findings were in general agreement with experiments, giving credibility to the report. The work predicted few important regions of the protein that can be mutated for improving the specificity of Cas9 and also offered new dynamic insights into allosterically triggered structural rearrangements of proximal and distal domains of the protein upon sgRNA and DNA binding.

Regarding the resolution of available crystal structures, cryogenic electron microscopy (cryo-EM) can certainly improve the quality of starting MD conformations. At present, there are only three entries corresponding to "CRIPSR/Cas9 in complex with sgRNA and DNA" available in the RCSB-PDB database that are obtained from cryo-EM (44). The cryo-EM atomic structure has been resolved for binary CRISPR/Cas complexes (45–47).

The paper by Palermo *et. al.* (48) should certainly be considered pioneering for the computational treatment of CRISPR/Cas9 complexes. The overall process of DNA editing by the CRISPR/Cas9 technique was already experimentally established. But the multi-microsecond all-atom MD simulations by Palermo *et. al.* revealed the conformational plasticity of the Cas9 endonuclease and identified the protein conformational changes that occur during nucleic acid binding and processing (48). The work specifically pointed towards the role of the ntDNA strand in the process of activation of the HNH domain for tDNA strand cleavage. MD simulations were carried out for four different systems: (i) apo-Cas9, (ii) a RNA/Cas9 binary complex, (iii) a RNA/DNA/Cas9 ternary complex in an intermediate state from PDB ID 4UN3 (27) and (iv) a RNA/DNA/Cas9 ternary complex in a pre-catalytic state from PDB ID 5F9R (28). The results of these computational experiments showed that the catalytic site H840 of the HNH domain is found at a distance of 25 Å from the scissile phosphate of the tDNA strand in the absence of ntDNA, incompatible with cleavage activity. The same work reported that instead, in the presence of ntDNA, H840 stabilizes at a smaller distance of 15 Å from the scissile phosphate of the tDNA strand **(Figure 5)**. It may seem that the distance of 15 Å is still too large for catalytic activity. Still, the authors pointed out that in the initial structure the same distance was 18 Å; hence, during the MD run the distance decreased to 15 Å in the presence of the ntDNA strand. This investigation was followed by the work by Zheng *et. al.* (43) based on coarse-grained modeling and "essential dynamics". The latter work confirmed the large-scale conformational transitions of Cas9 from the unbound form to the binary Cas9-RNA complex and then to the tertiary Cas9-RNA-DNA complex.

Before publication of the work by Palermo *et. al.* (48), the scientific community mostly believed that, since the HNH domain lies between the RuvC II and RuvC III motifs, it should be minimally involved in interactions with the rest of the protein. The computational data of Palermo *et. al.* (48), together with experimental data published in 2016 by Slaymaker *et. al.* (29), confirmed the importance of the HNH domain in the concerted editing mechanism. Moreover, according to Slaymaker *et. al.* (29), the u-ntDNA strand can be accommodated at the HNH/RuvC interface, through stabilizing electrostatic interactions between the negatively charged backbone of u-ntDNA and the positively charged amino acids at the HNH/RuvC interface. This evidence was exploited to build a model CRISPR/Cas9 ternary complex with a significantly long u-ntDNA strand (31), which was reported in 2017 and is discussed later in this review.

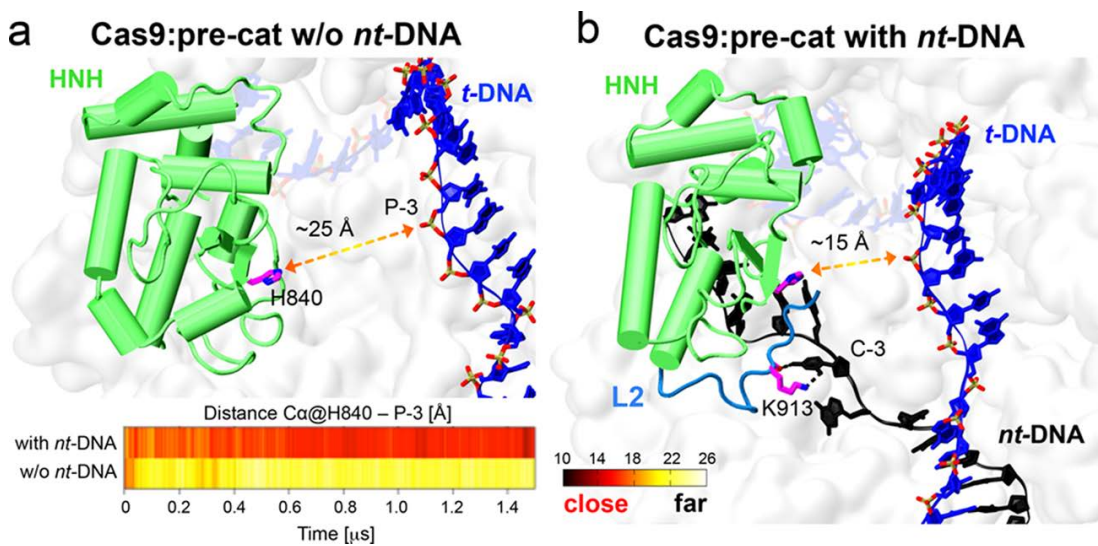

Figure 5: Representative structures from the MD trajectory obtained starting from the ternary complex in PDB ID 5F9R, (a) without the ntDNA strand (w/o nt-DNA) and (b) with the ntDNA strand. In the absence of the ntDNA strand, the catalytic H840 moves to a distance of ~25 Å from the scissile phosphate P-3 on the tDNA strand, being initially at ~18 Å distance. In the presence of the ntDNA strand, H840 approaches P-3 at ~15 Å distance. From Ref. (48), with permission.

As observed in the crystal structure (28), the HNH and RuvC domains are linked by hinge regions LI (residues 765-780) and LII (residues 906-918) **(Figure 4)** with an extended network of H-bonds. The LI-ntDNA contacts observed in the crystal structure are well conserved during the MD simulations (48). More importantly, the K913 residue of the LII loop forms an H-bond with a nucleotide of the ntDNA strand, which is stable over the time scale of ~0.75 ns during the MD run. This conformation is suggested to stabilize the H840 catalytic site of HNH domain at a distance of ~15 Å from the scissile phosphate on the tDNA strand, and cleavage of the tDNA strand by H840 follows. The communication between the RuvC domain and the ntDNA strand that influences the HNH catalytic domain positioning has been termed "crosstalk" between HNH and RuvC domains, which is dependent on the interactions between the hinge region and ntDNA strand (48). Thus, we now believe that the arrangement of the u-ntDNA strand in the complex is crucial for DNA editing. Another important aspect that was described in the same work (48) is that, in order for the ntDNA to bind the RuvC domain, it is important for the HNH domain to reposition, which is possible because the HNH domain has an inherent conformational plasticity, as revealed from comparative MD simulations of CRISPR/Cas9 ternary complexes in pre-catalytic (PDB ID 5F9R) and presumably post-catalytic (PDB ID 4UN3) states.

The work by Palermo *et. al*. published in 2017 in the Proceedings of the National Academy of Sciences USA (30) is devoted to the recognition and binding of the RNA by Cas9, poorly addressed previously. The computational results highlighted two significant facts: (i) conformational changes occur during CRISPR/Cas9 activation, specifically characterizing the conformational transition of Cas9 from its apo form to the RNA-bound form, thereby suggesting a mechanism for RNA binding within the protein; and (ii) crucial conformational changes in the HNH domain lead to a catalytically competent Cas9 conformation ready for cleaving the DNA. A dramatic conformational change was observed between the apo and RNA-bound structures of Cas9 by targeted MD (TMD) simulations. The apo Cas9 was obtained by removing the RNA from the RNA/Cas9 complex. The TMD trajectories revealed conformational rearrangements that mainly involve the REC lobe of Cas9. The REC lobe was observed to open up with respect to the comparatively stable nuclease lobe. During this process, the arginine rich helix was exposed into the solvent, forming a positively charged cavity that can accommodate the RNA. Single-molecule Foster Resonance Energy Transfer (FRET) experiments could reveal insights into the mechanism of RNA recognition at the atomistic level. In the simulations, the three regions RECI, RECII and RECIII that form the REC lobe were observed to move in different directions. In particular, the RECI region, together with the arginine rich helix, moves in the opposite direction as compared to the RECIII region. These observations are consistent with previous experimental (49) and computational studies (43, 48), suggesting that protein domains move concertedly in different directions for nucleic acid recognition. Palermo *et. al.* (30) compared the Cas9 conformational changes to an "earth and moon" model **(Figure 6),** where each individual protein domain rotates around the main protein axis ($\theta$) and itself ($\phi$). The extent of rotation of RECIII (theta angle of ~60° with respect to the protein and phi angle of ~90° with respect to itself) was much larger than the rotation of RECII.

The MD trajectories indicate that the high flexibility of the HNH domain controls the DNA cleavage, which is in agreement with biochemical evidences (50). As already mentioned, this flexibility depends on the conformational dynamics of the linker region between the HNH and RuvC domains. Further evidence for the role of linkers is provided by Palermo

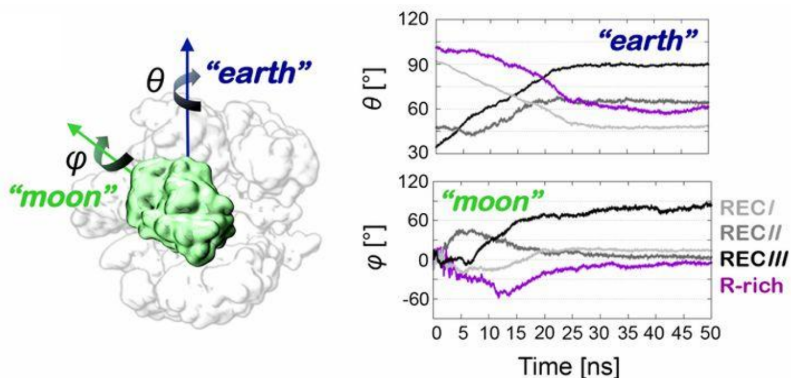

Figure 6: Schematic view of an earth and moon model of the system, where the θ and φ angles describe the rotation of each individual protein domain (here shown for HNH) with respect to the protein (θ) and itself (φ). The graphs report the time evolution of the θ ("earth") and φ ("moon") angles for the regions RECI-III and for the R-rich helix, along targeted MD. Adapted from Ref. (30), with permission.

*et. al.* (30) by TMD simulations and confirmed by single-molecule FRET experiments by Dagdas *et al.* (51). In a nutshell, it is proposed that collaborative interactions between LI/LII linkers and DNA cause the activation of RuvC and HNH domains for concerted cleavage of the DNA strands.

The active and inactive states of HNH were well distinguished by Gaussian-accelerated MD (GaMD) simulations (30), which reveal that the transition of the HNH domain from the inactive to the active state should involve a critical step, namely binding of dsDNA. This evidence was supported by the TMD simulations of the HNH transition (30), suggesting that HNH repositioning might occur during DNA unwinding, leading to the speculation that the high flexibility of the HNH domain might somehow facilitate unwinding of dsDNA and formation of R-loop structure (48–50). Overall, by using equilibrium MD, GaMD and TMD simulations, the conformational dynamics involved in RNA recognition, DNA unwinding and DNA cleavage by Cas9 has been substantially explained.

The first step in CRISPR/Cas9 genome editing is, however, recognition of the PAM motif: the mechanism through which PAM binding activates Cas9 for DNA cleavage at spatially distant sites still needs understanding. Palermo *et. al.* (52) provided evidence of a PAM-induced allosteric mechanism in CRISPR/Cas9 by MD sampling over ~13μs, employing simulation conditions tailored for DNA/RNA endonucleases, for two different systems: Cas9 with guide RNA and its matching DNA containing a 5′-TGG-3′ PAM sequence (referred as Cas9 with PAM: Cas9-wPAM, from PDB ID 4UN3 (27)),  and its analogue, crystallized without the PAM segment (Cas9-w/oPAM, from PDB ID 4OO8 (7)). The work concluded that PAM functions as an allosteric effector that triggers interdependent conformational dynamics of the HNH and RuvC domains, required for concerted cleavage of DNA strands. A similar allosteric communication was observed using intramolecular FRET experiments by Sternberg *et. al.* (50).

Principal modes of motion obtained from MD simulations of Cas9-wPAM and Cas9-w/oPAM revealed an "open-to-close" conformational transition for Cas9, which is necessary for nucleic acid binding (30, 48). In the presence of PAM, the correlated residues exhibit higher correlation strength necessary for allosteric signaling, as compared to the correlation strength in the absence of PAM. The PAM-induced conformational transition increases the strength of correlated motions between RuvC and HNH domains, inducing a stronger communication channel between these two nuclease domains. Allosteric signaling is triggered by electrostatic interactions between charged amino acids (16). Palermo *et. al.* specifically pointed out that Q771 and E584 interact with K775 and R905, respectively, forming essential edges of the allosteric pathway. These interactions connect HNH to RuvC and the α-helical lobe via the linkers LI and LII, which effectively function as "allosteric transducers" (28, 50). The optimal path for information flow was identified through the LI loop, via residues K772 and T770. Very recently, the allosteric mechanism of information transfer across the HNH domain has been investigated by East *et. al.* (53) using a combination of solution NMR experiments and MD simulations. The complete allosteric pathway spans the HNH domain from the region interfacing the RuvC domain up to the RECII region. The DNA binding signal propagates across the recognition lobe and the nuclease domains (HNH and RuvC), for concerted cleavage of the t-DNA and nt-DNA strands.

Molecular simulations so far have significantly contributed to understanding Cas9 structure and function, as summarized above. Yet, little is known about the catalytically active states of Cas9 HNH and RuvC domains. The first model of the fully catalytic CRISPR/Cas9 was reported by Palermo *et. al*. in 2017 in the Proceedings of the National Academy of Sciences USA (30), already cited above. In 2016, Zuo and Liu reported the catalytically competent state of the RuvC domain by MD simulations (35), but the HNH conformations were elusive. In 2017 (54), the same authors reported the missing link to decipher how the HNH domain transitions from the pre-catalytic state to the catalytic state. They employed two distinct sampling techniques: biased targeted-MD (tMD) and unbiased ensemble conventional MD (cMD$^{ens}$). They obtained a cross-validated catalytically active state of the Cas9 HNH domain primed for cutting the tDNA strand. The *Sp*Cas9 crystal structures corresponding to different binding stages (6, 7, 27, 28, 49) can be used to understand the conformational activation pathway of the HNH domain of Cas9 **(Figure 7)**. Zuo and Liu found that $Mg^{2+}$ ions are indispensable for the formation and stability of the catalytic state of the HNH domain, which is in accordance with experimental reports by Jinek *et. al.* (4) and Dagdas *et. al.* (51). In addition, the results by Zuo and Liu suggest that $Mg^{2+}$ ions also act as facilitators and stabilizers of the functional conformational state.

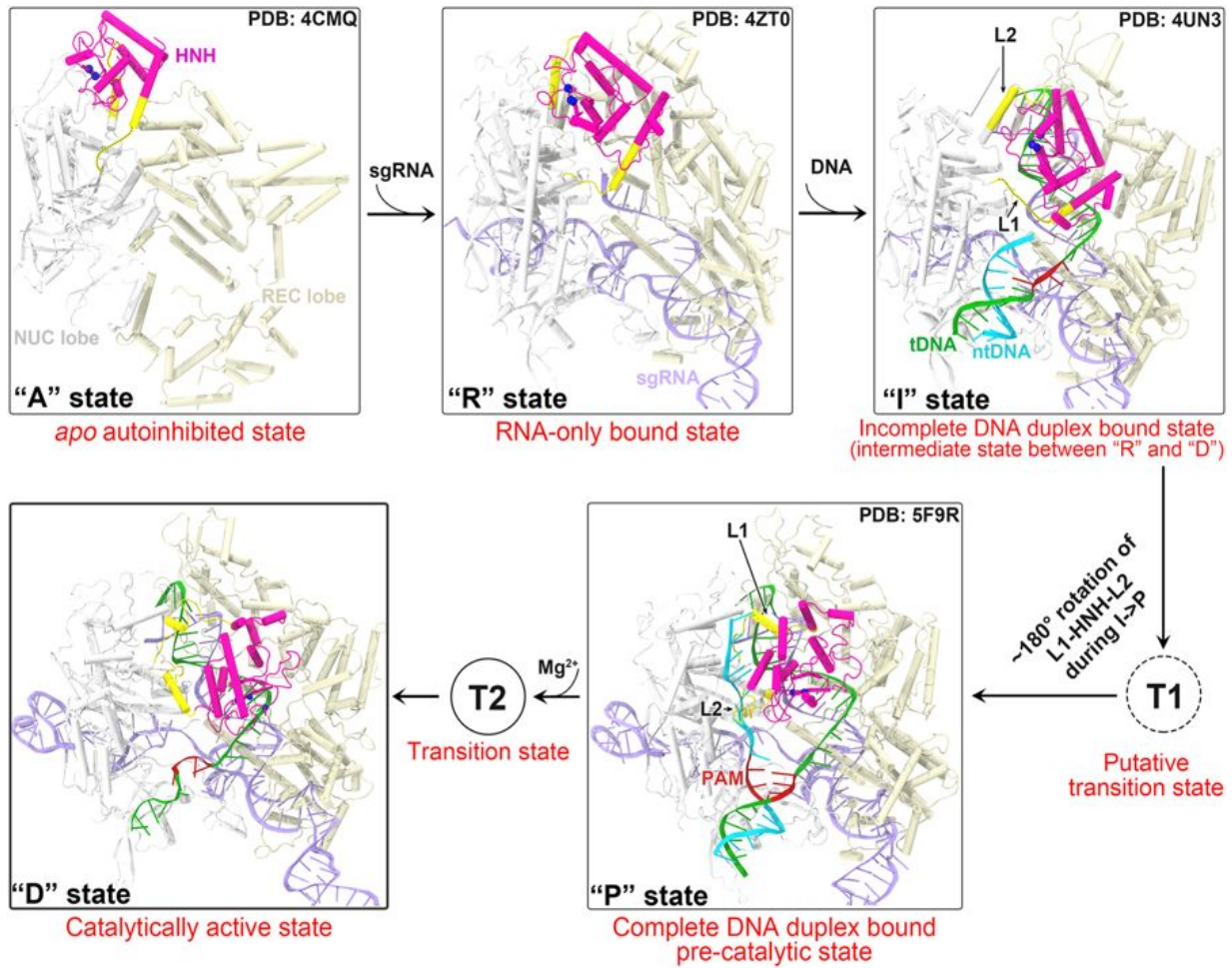

Figure 7: Conformational activation pathway of Cas9 HNH nuclease domain. The HNH domain and flanking linker regions (i.e. L1 and L2) are highlighted in magenta and yellow, respectively. The PAM is colored dark red and the three putative catalytic residues of HNH domain are represented as blue spheres. From Ref. (54) with permission.

Understanding the catalytic mechanism involved in accurate editing by CRISPR/Cas9 is particularly important, but it has been poorly addressed so far, due to uncertainties in the available crystal structures of Cas9. Recently, Yoon *et. al.* (55) attempted to analyze the catalytic mechanism and the energetics for the activation of Cas9, by using the empirical valence-bond (EVB) method. However, the investigation was based on a Cas9 analogous protein, the endonuclease VII and the conclusions were made based on the structural similarity between Cas9 and endonuclease VII. Yoon and co-workers specifically looked for the catalytic roles of two positive residues K848 and K855 in the HNH domain. They found that the catalytic state most likely involves a structure where K848 comes close to the scissile phosphate of tDNA strand.

Little is known about the catalytic state of the nuclease domain in CRISPR/Cas9 and even less is known about the location and binding of divalent metal ions within the RuvC nuclease domain active site, necessary for cleavage of nt-DNA strand. Zuo and Liu in 2016 (35) attempted to understand the metal-aided cleavage mechanism by classical MD simulations. The identified catalytic site shows that the pro-Rp and pro-Sp oxygen atoms of the scissile phosphate coordinate the MgA and MgB ions, respectively **(Figure 8a)**. This is unlikely to give rise to phosphodiester bond cleavage, which instead would require only the pro-Sp oxygen of the scissile phosphate to jointly coordinate with the two $Mg^{2+}$ ions, enabling an in-line SN2 nucleophilic attack. This is a well-known requirement for the two-metal aided catalysis (36, 56). Using QM/MM simulations with the quantum mechanical portion of the system described at the DFT level, Palermo and co-authors revealed in 2019 a new geometry for the active site with joint coordination of the two $Mg^{2+}$ ions by the pro-Sp oxygen of the scissile phosphate **(Figure 8b)** (40).

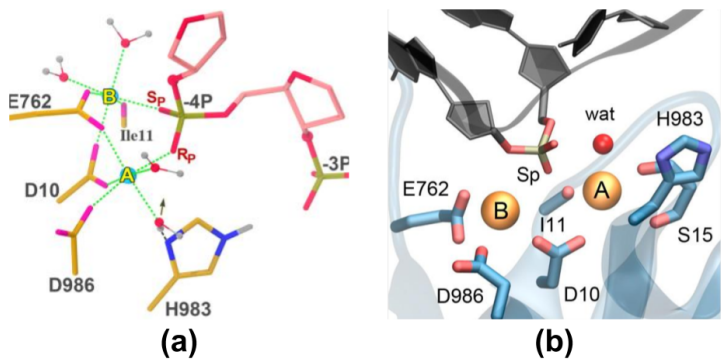

Figure 8: The active site geometry of RuvC domain as obtained from (a) classical MD by Zuo and Liu in 2016 and (b) QM/MM by Palermo *et. al.* in 2019. From Ref 35 and Ref 40, respectively, with permissions.

This agrees well with the requirements for the catalysis, as it favors an in-line SN2-like nucleophilic attack. Thus, it is likely that the active site geometry obtained by Zuo and Liu via classical MD might have been affected by the limitations of the classical force field for $Mg^{2+}$ ions and the same limitation may be responsible for "staggered cleavage".

The computational works discussed above were dedicated to understanding the structure, dynamics and function of the NUC lobe of Cas9 (or other analogous protein), specifically the nuclease domains HNH and RuvC. Other parts of the protein may participate in structural transformations to attain cleavage-prone conformations. A recently published work by Palermo *et. al.* (57), based on 16 μs of unbiased MD simulations performed using Anton-2, revealed that the conformational change in the HNH domain during DNA cleavage is accompanied by a remarkable structural remodeling of the REC lobe. The conformational changes in the REC lobe indicate that the RECI, RECII and RECIII domains 'sense' nucleic acids, 'regulate' the HNH conformational transition and 'lock' the HNH domain at the cleavage site, leading to the formation of a catalytically active CRISPR/Cas9 complex. Thus, a subtle interplay between the REC and NUC lobes is observed upon activation, ultimately ensuring catalytic competence.

The CRISPR/Cas9 approach has been exploited for efficient genome editing in a wide variety of organisms. However, the method exhibits a major undesired effect: it results in unwanted editing at off-target sites that are similar to on-target DNA sequences (58, 59). It is important to make progress towards eliminating such off-target effects on a genome-wide scale. Within the framework of the hypothesis proposed for Cas9-sgRNA (29, 58), the structural information obtained from all previous MD investigations has been exploited to rationally design Cas9 variants with improved specificity. Palermo and coworkers (60) devoted attention towards off-target cleavage. The authors decipher the mechanism of off-target binding at the molecular level, using GaMD simulations. They conclude that base pair mismatches in the target DNA at specific distal sites with respect to the PAM induce an extended opening of the RNA-DNA hybrid duplex, which leads to newly discovered interactions between the unwound nucleic acids and the protein counterpart. This work, which focuses on interactions between DNA and the protein HNH domain, poses the foundations for designing novel and more specific Cas9 variants. Based on the structural information obtained from their MD simulations, Zuo and Liu suggested more than a dozen potential mutation sites in the NUC lobe of Cas9 for generating Cas9 variants with improved specificity (54). On the other hand, Palermo *et. al.* (57) focused on the REC lobe to control off-target effects.

In addition to dynamics of wild-type *Sp*Cas9, the dynamics of Cas9 mutants have also been explored using MD simulations for understanding off-target effects. Zheng *et. al.* (61) selected four mutants: N497A, R661A, Q695A and Q926A (referred to as *nrqq*Cas9). This choice was based on the experimental work by Kleinstiver *et. al.* (58), that show ability to decrease all, or nearly all, genome-wide off-target effects. The work by Zheng and co-workers (56) provided the much-needed atomistic explanation for the lowering of off-target effects accompanied by lowering of on-target cleavage efficiency of *nrqq*Cas9. The results indicated that the mutations cause a loss in the electrostatic interactions between Cas9 and the RNA-DNA hybrid duplex, especially in the PAM distal segment. The *nrqq*Cas9 systems adopt slightly open conformation with fewer interactions between HNH and RuvC domains, with consequently reduced efficiency for on-target and off-target DNA cleavage.

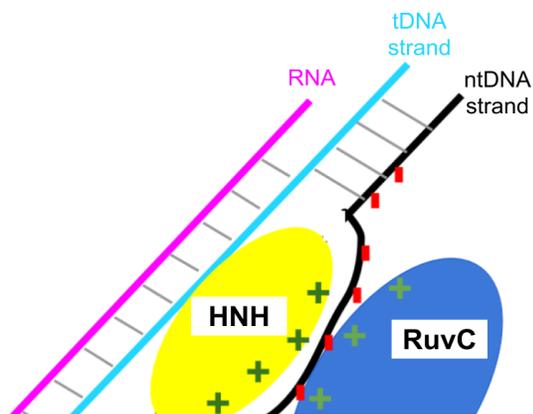

Figure 9: Scheme of various competitive interactions in the system.protein-ntDNA, t-DNA-ntDNA, tDNA-RNA. If the protein-ntDNA interaction strength is decreased by mutating charged Cas9 residues to neutral residues, there will be a driving force towards DNA duplex re-hybridization, unless the hybrid RNA:tDNA duplex is very strong, which is the case for perfect complementarity.

The issue of specificity is also currently addressed in our group, by MD simulations of CRISPR/Cas9 complexes with protein mutations at the HNH/RuvC boundary. Our work is inspired by the suggestion, based on experimental data (29), that attractive electrostatic interactions between positive protein residues at this boundary and the negative backbone of the u-ntDNA strand keep the u-ntDNA strand unwounded **(Figure 9)**, thus preventing the re-formation of the DNA duplex and keeping the system ready for cleavage. Protein mutations from charged to neutral amino acids can weaken the protein-ntDNA attraction and induce the duplex DNA re-hybridization if the RNA:tDNA duplex is not strong enough, e.g. in the case of imperfect complementarity. We have already positively tested this model **(Figure 10)** using a 1.5 μs MD trajectory of a modeled structure derived from PDB ID 4UN3, with an elongated ntDNA strand accommodated at the HNH/RuvC positive patch (31). We validated this model by a combination of MD and electron paramagnetic (EPR) measurements of spin-labeled CRISPR/Cas9 ternary complexes, with nitroxide spin labels on the DNA backbone. EPR measured distances between chosen DNA sites resulted in agreement with distances between the same sites extracted from the MD trajectory with nitroxides added through the software NASNOX (62), which accounts for all sterically allowed rotamers.

Until now, MD simulations have been mostly utilized to characterize the interactions between different segments of CRISPR/Cas9: Cas9-tDNA, Cas9-ntDNA, PAM-DNA, etc.

Attention has been mostly focused on the HNH and RuvC domains, as these compose the catalytic region and contain the nuclease active site, with very little work on the REC lobe. We anticipate the need for large amounts of experimental and theoretical/computational work to explore other parts of the protein, which may play crucial allosteric roles in nucleic acid recognition and cleavage. More extensive comprehension of structural design will eventually enable scientists to tailor specific Cas9 variants with no off-target effects.

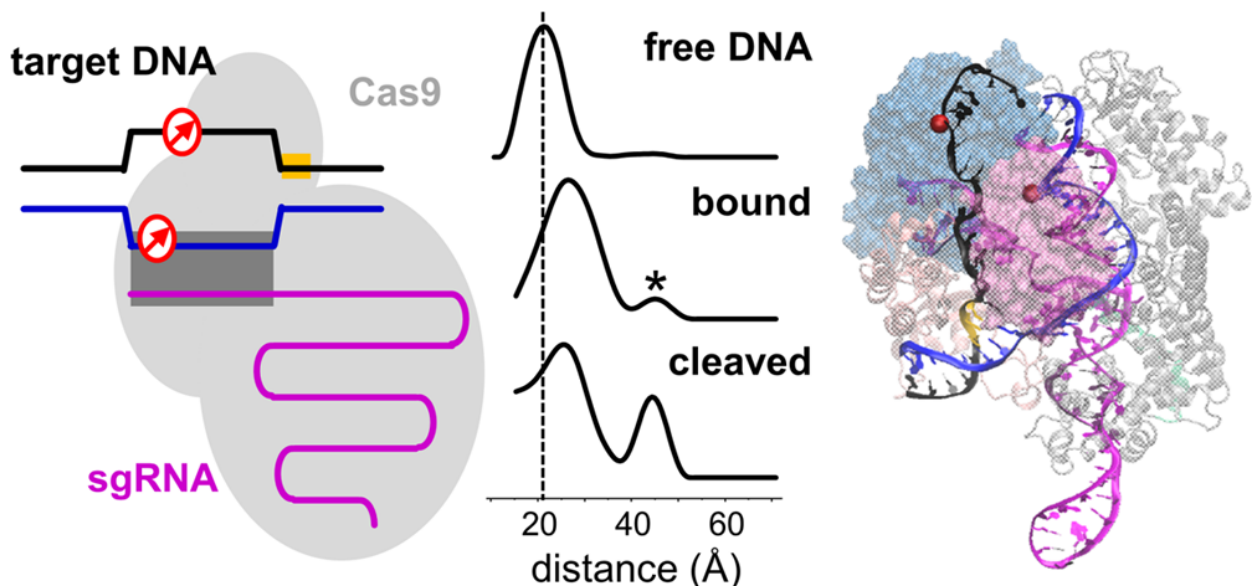

Figure 10: Left: scheme of the experimental system for EPR measurements. Center: Measured distance profiles for selected DNA sites. Right: simulated structure, with the elongated ntDNA in black and the sites for distance calculations visualized as spheres. From Ref (31) with permission.

**Limitations and Future Directions**

The nuclease activity of Cas9 can be triggered even when there is imperfect complementarity between the sgRNA and an off-target genomic site, particularly if mismatches are distal to PAM (27, 59, 63). These off-target effects are serious limitations to the practical use of CRISPR/Cas in genetic therapies, agriculture and other potential applications with societal impact. Anders *et. al.* (27). crystallized two Cas9/sgRNA/DNA complexes, where the DNA contains mismatches to the guide RNA. They revealed that, even in the absence of full 20-nt complementary base pairing between DNA and RNA, target DNA binding by Cas9/RNA results in local strand separation immediately upstream of the PAM, thereby leading to cleavage of DNA strands in undesired conditions. Such off-target activities are prominent when the mismatches are distal to PAM. Slaymaker *et. al.* (29) reported a structure-guided protein engineering method that promises to improve the specificity of the CRISPR technology, at least when the endonuclease is *Sp*Cas9. It is hypothesized that neutralizing the positively charged residues at the interface between HNH and RuvC domains, where the u-ntDNA strand is hosted, could weaken the ntDNA strand binding to protein **(Figure 9)**. This would encourage re-hydridization between the tDNA and ntDNA strands, causing "R-loop collapse". Hence, a more stringent Watson-Crick base pairing between guide RNA and target DNA would be necessary to prevent the R-loop collapse. Slaymaker *et. al.* (29) initially generated thirty-one *Sp*Cas9 mutants replacing charged amino acids at the HNH/RuvC interface with alanine. The top five specificity conferring mutants were R780A, K810A, K848A, K855A and H982A. We are currently performing μs-long MD simulations to investigate the structural and dynamical basis of this intriguing hypothesis. Specifically, we are addressing CRISPR/Cas9 ternary complexes with wild-type Cas9 and three Cas9 mutants: K855A, H982A and a combination K855A+H982A. We use the system derived from PDB ID 4UN3, with an elongated u-ntDNA strand (**Figure 9**) (31) and modeled missing Cas9 residues. Preliminary results (64), obtained by the comparison between complexes with wild-type and mutated Cas9 complexes, yield structural insights into the active DNA-bound state of the RuvC domain, highlighting motions that are consistent with the competing electrostatic interactions as illustrated in **Figure 9**. Our preliminary results concur with the hypothesis that protein mutations can be exploited to produce more specific CRISPR/Cas9 variants.

A sgRNA, formed by crRNA and tracrRNA, and the PAM region in the DNA, are essential for Cas9-mediated dsDNA cleavage in the presence of $Mg^{2+}$ ions. Sundaresan *et. al.* (65), however, demonstrated that FnoCas9 and FnoCas12a possess RNA-independent, non-sequence-specific cleavage activities on dsDNA targets in the presence of $Mn^{2+}$. This finding emphasizes the need for further characterization of CRISPR systems towards the development of applications to living genomes.

**References**

1. Baig, A.M. 2018. Human Genome-Edited Babies: First Responder with Concerns Regarding Possible Neurological Deficits! ACS Chem. Neurosci. 10: 8b00668.
2. Consortium, I.H.G.S. 2001. Initial sequencing and analysis of the human genome. Nature. 409: 860–921.
3. Ishino, Y., H. Shinagawa, K. Makino, M. Amemura, and A. Nakata. 1987. Nucleotide sequence of the iap gene, responsible for alkaline phosphatase isozyme conversion in Escherichia coli, and identification of the gene product. J. Bacteriol. 169: 5429–33.
4. Jinek, M., K. Chylinski, I. Fonfara, M. Hauer, J.A. Doudna, and E. Charpentier. 2012. A programmable dual-RNA-guided DNA endonuclease in adaptive bacterial immunity. Science. 337: 816–21.
5. Cong, L., F.A. Ran, D. Cox, S. Lin, R. Barretto, N. Habib, P.D. Hsu, X. Wu, W. Jiang, L.A. Marraffini, and F. Zhang. 2013. Multiplex genome engineering using CRISPR/Cas systems. Science. 339: 819–23.
6. Jinek, M., F. Jiang, D.W. Taylor, S.H. Sternberg, E. Kaya, E. Ma, C. Anders, M. Hauer, K. Zhou, S. Lin, M. Kaplan, A.T. Iavarone, E. Charpentier, E. Nogales, and J.A. Doudna. 2014. Structures of Cas9 Endonucleases Reveal RNA-Mediated Conformational Activation. Science. 343: 1247997–1247997.
7. Nishimasu, H., F.A. Ran, P.D. Hsu, S. Konermann, S.I. Shehata, N. Dohmae, R. Ishitani, F. Zhang, and O. Nureki. 2014. Crystal Structure of Cas9 in Complex with Guide RNA and Target DNA. Cell. 156: 935–949.
8. Karginov, F. V, and G.J. Hannon. 2010. The CRISPR system: small RNA-guided defense in bacteria and archaea. Mol. Cell. 37: 7–19.
9. Rath, D., L. Amlinger, A. Rath, and M. Lundgren. 2015. The CRISPR-Cas immune system: Biology, mechanisms and applications. Biochimie. 117: 119–128.
10. Mohanraju, P., K.S. Makarova, B. Zetsche, F. Zhang, E. V. Koonin, and J. van der Oost. 2016. Diverse evolutionary roots and mechanistic variations of the CRISPR-Cas systems. Science. 353: aad5147.
11. Thurtle-Schmidt, D.M., and T.-W. Lo. 2018. Molecular biology at the cutting edge: A review on CRISPR/CAS9 gene editing for undergraduates. Biochem. Mol. Biol. Educ. 46: 195–205.
12. Makarova, K.S., D.H. Haft, R. Barrangou, S.J.J. Brouns, E. Charpentier, P. Horvath, S. Moineau, F.J.M. Mojica, Y.I. Wolf, A.F. Yakunin, J. van der Oost, and E. V Koonin. 2011. Evolution and classification of the CRISPR-Cas systems. Nat. Rev. Microbiol. 9: 467–77.
13. Shmakov, S., A. Smargon, D. Scott, D. Cox, N. Pyzocha, W. Yan, O.O. Abudayyeh, J.S. Gootenberg, K.S. Makarova, Y.I. Wolf, K. Severinov, F. Zhang, and E. V. Koonin. 2017. Diversity and evolution of class 2 CRISPR–Cas systems. Nat. Rev. Microbiol. 15: 169–182.
14. Koonin, E. V, K.S. Makarova, and F. Zhang. 2017. Diversity, classification and evolution of CRISPR-Cas systems. Curr. Opin. Microbiol. 37: 67–78.
15. Swarts, D.C., and M. Jinek. 2018. Cas9 versus Cas12a/Cpf1: Structure-function comparisons and implications for genome editing. Wiley Interdiscip. Rev. RNA. 9: e1481.
16. Makarova, K.S., Y.I. Wolf, O.S. Alkhnbashi, F. Costa, S.A. Shah, S.J. Saunders, R. Barrangou, S.J.J. Brouns, E. Charpentier, D.H. Haft, P. Horvath, S. Moineau, F.J.M. Mojica, R.M. Terns, M.P. Terns, M.F. White, A.F. Yakunin, R.A. Garrett, J. van der Oost, R. Backofen, and E. V. Koonin. 2015. An updated evolutionary classification of CRISPR–Cas systems. Nat. Rev. Microbiol. 13: 722–736.
17. Makarova, K.S., Y.I. Wolf, and E. V. Koonin. 2018. Classification and Nomenclature of

CRISPR-Cas Systems: Where from Here? Cris. J. 1: 325–336.
18. Jiang, F., and J.A. Doudna. 2017. CRISPR–Cas9 Structures and Mechanisms. Annu. Rev. Biophys. 46: 505–529.
19. Sternberg, S.H., S. Redding, M. Jinek, E.C. Greene, and J.A. Doudna. 2014. DNA interrogation by the CRISPR RNA-guided endonuclease Cas9. Nature. 507: 62–67.
20. Gong, S., H.H. Yu, K.A. Johnson, and D.W. Taylor. 2018. DNA Unwinding Is the Primary Determinant of CRISPR-Cas9 Activity. Cell Rep. 22: 359–371.
21. Deltcheva, E., K. Chylinski, C.M. Sharma, K. Gonzales, Y. Chao, Z.A. Pirzada, M.R. Eckert, J. Vogel, and E. Charpentier. 2011. CRISPR RNA maturation by trans-encoded small RNA and host factor RNase III. Nature. 471: 602–607.
22. Jinek, M., K. Chylinski, I. Fonfara, M. Hauer, J.A. Doudna, and E. Charpentier. 2012. A Programmable Dual-RNA–Guided DNA Endonuclease in Adaptive Bacterial Immunity. Science. 337: 816–821.
23. Hille, F., and E. Charpentier. 2016. CRISPR-Cas: biology, mechanisms and relevance. Philos. Trans. R. Soc. Lond. B. Biol. Sci. 371.
24. Rueda, F.O., M. Bista, M.D. Newton, A.U. Goeppert, M.E. Cuomo, E. Gordon, F. Kröner, J.A. Read, J.D. Wrigley, D. Rueda, and B.J.M. Taylor. 2017. Mapping the sugar dependency for rational generation of a DNA-RNA hybrid-guided Cas9 endonuclease. Nat. Commun. 8: 1610.
25. CRISPR/Cas9 & Targeted Genome Editing: New Era in Molecular Biology. https://www.neb.com/tools-and-resources/feature-articles/crispr-cas9-and-targeted-genome-editing-a-new-era-in-molecular-biology
26. Lee, J., J.-H. Chung, H.M. Kim, D.-W. Kim, and H. Kim. 2016. Designed nucleases for targeted genome editing. Plant Biotechnol. J. 14: 448–462.
27. Anders, C., O. Niewoehner, A. Duerst, and M. Jinek. 2014. Structural basis of PAM-dependent target DNA recognition by the Cas9 endonuclease. Nature. 513: 569–573.
28. Jiang, F., D.W. Taylor, J.S. Chen, J.E. Kornfeld, K. Zhou, A.J. Thompson, E. Nogales, and J.A. Doudna. 2016. Structures of a CRISPR-Cas9 R-loop complex primed for DNA cleavage. Science. 351: 867–871.
29. Slaymaker, I.M., L. Gao, B. Zetsche, D.A. Scott, W.X. Yan, and F. Zhang. 2016. Rationally engineered Cas9 nucleases with improved specificity. Science. 351: 84–88.
30. Palermo, G., Y. Miao, R.C. Walker, M. Jinek, and J.A. McCammon. 2017. CRISPR-Cas9 conformational activation as elucidated from enhanced molecular simulations. Proc. Natl. Acad. Sci. U. S. A. 114: 7260–7265.
31. Tangprasertchai, N.S., R. Di Felice, X. Zhang, I.M. Slaymaker, C. Vazquez Reyes, W. Jiang, R. Rohs, and P.Z. Qin. 2017. CRISPR–Cas9 Mediated DNA Unwinding Detected Using Site-Directed Spin Labeling. ACS Chem. Biol. 12: 1489–1493.
32. Humphrey, W., A. Dalke, and K. Schulten. 1996. VMD: visual molecular dynamics. J. Mol. Graph. 14: 33–8, 27–8.
33. Jinek, M., A. East, A. Cheng, S. Lin, E. Ma, and J. Doudna. 2013. RNA-programmed genome editing in human cells. 2: 471.
34. Mali, P., L. Yang, K.M. Esvelt, J. Aach, M. Guell, J.E. DiCarlo, J.E. Norville, and G.M. Church. 2013. RNA-guided human genome engineering via Cas9. Science. 339: 823-826
35. Zuo, Z., and J. Liu. 2016. Cas9-catalyzed DNA Cleavage Generates Staggered Ends: Evidence from Molecular Dynamics Simulations. Sci. Rep. 6: 37584.
36. Yang, W. 2011. Nucleases: diversity of structure, function and mechanism. Q. Rev.

Biophys. 44: 1–93.
37. Yang, W. 2008. An equivalent metal ion in one- and two-metal-ion catalysis. Nat. Struct. Mol. Biol. 15: 1228–1231.
38. Yang, W., J.Y. Lee, and M. Nowotny. 2006. Making and Breaking Nucleic Acids: Two-Mg2+-Ion Catalysis and Substrate Specificity. Mol. Cell. 15: 1228–123.
39. Li, P., and K.M. Merz. 2017. Metal Ion Modeling Using Classical Mechanics. Chem. Rev. 117: 1564–1686.
40. Palermo, G. 2019. Structure and Dynamics of the CRISPR–Cas9 Catalytic Complex. J. Chem. Inf. Model. 59: 2394–2406.
41. Estarellas, C., M. Otyepka, J. Koča, P. Banáš, M. Krepl, and J. Šponer. 2015. Molecular dynamic simulations of protein/RNA complexes: CRISPR/Csy4 endoribonuclease. Biochim. Biophys. Acta - Gen. Subj. 1850: 1072–1090.
42. Shaw, D.E., J.P. Grossman, J.A. Bank, B. Batson, J.A. Butts, J.C. Chao, M.M. Deneroff, R.O. Dror, A. Even, C.H. Fenton, A. Forte, J. Gagliardo, G. Gill, B. Greskamp, C.R. Ho, D.J. Ierardi, L. Iserovich, J.S. Kuskin, R.H. Larson, T. Layman, L.-S. Lee, A.K. Lerer, C. Li, D. Killebrew, K.M. Mackenzie, S.Y.-H. Mok, M.A. Moraes, R. Mueller, L.J. Nociolo, J.L. Peticolas, T. Quan, D. Ramot, J.K. Salmon, D.P. Scarpazza, U. Ben Schafer, N. Siddique, C.W. Snyder, J. Spengler, P.T.P. Tang, M. Theobald, H. Toma, B. Towles, B. Vitale, S.C. Wang, and C. Young. 2014. Anton 2: Raising the Bar for Performance and Programmability in a Special-Purpose Molecular Dynamics Supercomputer. In: SC14: International Conference for High Performance Computing, Networking, Storage and Analysis. IEEE. pp. 41–53.
43. Zheng, W. 2017. Probing the structural dynamics of the CRISPR-Cas9 RNA-guided DNA-cleavage system by coarse-grained modeling. Proteins Struct. Funct. Bioinforma. 85: 342–353.
44. Zhu, X., R. Clarke, A.K. Puppala, S. Chittori, A. Merk, B.J. Merrill, M. Simonovic, and S. Subramaniam. 2019. Cryo-EM structures reveal coordinated domain motions that govern DNA cleavage by Cas9. Nat.Struct.Mol.Biol. 10.2210/PDB6O0X/PDB.
45. Guo, T.W., A. Bartesaghi, H. Yang, L.A. Earl, D.J. Patel, and G. Et Al. 2017. Cryo-EM Structures Reveal Mechanism and Inhibition of DNA Targeting by a CRISPR-Cas Surveillance Complex Data Resources 6B44 6B45 6B46 6B47 6B48. Cell. 171: 414–426.
46. Dorsey, B.W., L. Huang, and A. Mondragón. 2019. Structural organization of a Type III-A CRISPR effector subcomplex determined by X-ray crystallography and cryo-EM. Nucleic Acids Res. 47: 3765–3783.
47. Huo, Y., T. Li, N. Wang, Q. Dong, X. Wang, and T. Jiang. 2018. Cryo-EM structure of Type III-A CRISPR effector complex. Cell Res. 28: 1195–1197.
48. Palermo, G., Y. Miao, R.C. Walker, M. Jinek, and J.A. McCammon. 2016. Striking Plasticity of CRISPR-Cas9 and Key Role of Non-target DNA, as Revealed by Molecular Simulations. ACS Cent. Sci. 2: 756–763.
49. Jiang, F., K. Zhou, L. Ma, S. Gressel, and J.A. Doudna. 2015. STRUCTURAL BIOLOGY. A Cas9-guide RNA complex preorganized for target DNA recognition. Science. 348: 1477–81.
50. Sternberg, S.H., B. LaFrance, M. Kaplan, and J.A. Doudna. 2015. Conformational control of DNA target cleavage by CRISPR–Cas9. Nature. 527: 110–113.
51. Dagdas, Y.S., J.S. Chen, S.H. Sternberg, J.A. Doudna, and A. Yildiz. 2017. A conformational checkpoint between DNA binding and cleavage by CRISPR-Cas9. Sci.

Adv. 3: eaao0027.
52. Palermo, G., C.G. Ricci, A. Fernando, R. Basak, M. Jinek, I. Rivalta, V.S. Batista, and J.A. McCammon. 2017. Protospacer Adjacent Motif-Induced Allostery Activates CRISPR-Cas9. J. Am. Chem. Soc. 139: 16028–16031.
53. East, K.W., J.C. Newton, U.N. Morzan, A. Acharya, E. Skeens, G. Jogl, V.S. Batista, G. Palermo, and G.P. Lisi. 2019. Allosteric Motions of the CRISPR-Cas9 HNH Nuclease Probed by NMR and Molecular Dynamics. bioRxiv. : 660613.
54. Zuo, Z., and J. Liu. 2017. Structure and Dynamics of Cas9 HNH Domain Catalytic State. Sci. Rep. 7: 17271.
55. Yoon, H., L.N. Zhao, and A. Warshel. 2019. Exploring the Catalytic Mechanism of Cas9 Using Information Inferred from Endonuclease VII. ACS Catal. 9: 1329–1336.
56. Steitz, T.A., and J.A. Steitz. 1993. A general two-metal-ion mechanism for catalytic RNA. Proc. Natl. Acad. Sci. U. S. A. 90: 6498–502.
57. Palermo, G., J.S. Chen, C.G. Ricci, I. Rivalta, M. Jinek, V.S. Batista, J.A. Doudna, and J.A. McCammon. 2018. Key role of the REC lobe during CRISPR–Cas9 activation by 'sensing', 'regulating', and 'locking' the catalytic HNH domain. Q. Rev. Biophys. 51: e9.
58. Kleinstiver, B.P., V. Pattanayak, M.S. Prew, S.Q. Tsai, N.T. Nguyen, Z. Zheng, and J.K. Joung. 2016. High-fidelity CRISPR–Cas9 nucleases with no detectable genome-wide off-target effects. Nature. 529: 490–495.
59. Fu, Y., J.A. Foden, C. Khayter, M.L. Maeder, D. Reyon, J.K. Joung, and J.D. Sander. 2013. High-frequency off-target mutagenesis induced by CRISPR-Cas nucleases in human cells. Nat. Biotechnol. 31: 822–6.
60. Ricci, C.G., J.S. Chen, Y. Miao, M. Jinek, J.A. Doudna, J.A. McCammon, and G. Palermo. 2019. Deciphering Off-Target Effects in CRISPR-Cas9 through Accelerated Molecular Dynamics. ACS Cent. Sci. : acscentsci.9b00020.
61. Zheng, L., J. Shi, and Y. Mu. 2018. Dynamics changes of CRISPR-Cas9 systems induced by high fidelity mutations. Phys. Chem. Chem. Phys. 20: 27439–27448.
62. Qin, P.Z., I.S. Haworth, Q. Cai, A.K. Kusnetzow, G.P.G. Grant, E.A. Price, G.Z. Sowa, A. Popova, B. Herreros, and H. He. 2007. Measuring nanometer distances in nucleic acids using a sequence-independent nitroxide probe. Nat. Protoc. 2: 2354–2365.
63. Hsu, P.D., D.A. Scott, J.A. Weinstein, F.A. Ran, S. Konermann, V. Agarwala, Y. Li, E.J. Fine, X. Wu, O. Shalem, T.J. Cradick, L.A. Marraffini, G. Bao, and F. Zhang. 2013. DNA targeting specificity of RNA-guided Cas9 nucleases. Nat. Biotechnol. 31: 827–832.
64. Ray, A., and R. Di Felice. Unpublished. .
65. Sundaresan, R., H.P. Parameshwaran, S.D. Yogesha, M.W. Keilbarth, and R. Rajan. 2017. RNA-Independent DNA Cleavage Activities of Cas9 and Cas12a. Cell Rep. 21: 3728–3739.